\documentclass[pra,twocolumn,a4paper,showpacs,floatfix]{revtex4} 

\usepackage{graphicx} \newcommand{\figspace}{\vspace*{2ex}}
\usepackage{amsmath}

\begin{document}

\title{Magneto-transport in periodic and quasiperiodic arrays of
  mesoscopic rings}

\author{Arunava Chakrabarti} \affiliation{Department of Physics,
  University of Kalyani, Kalyani, West Bengal 741 235, India}
\author{Rudolf A.
  R\"{o}mer}\altaffiliation[Email:]{r.roemer@warwick.ac.uk}

\affiliation{Department of Physics and Centre for Scientific Computing,
  University of Warwick, Coventry CV4 7AL, United Kingdom}
\author{Michael Schreiber}
\affiliation{Institut f\"{u}r Physik, Technische Universit\"{a}t,
09107 Chemnitz, Germany}

\date{$Revision: 1.50 $, compiled \today}

\begin{abstract}
  We study theoretically the transmission properties of serially
  connected mesoscopic rings threaded by a magnetic flux. Within a
  tight-binding formalism we derive exact analytical results for the
  transmission through periodic and quasiperiodic Fibonacci arrays of
  rings of two different sizes. The role played by the number of
  scatterers in each arm of the ring is analyzed in some detail. The
  behavior of the transmission coefficient at a particular value of the
  energy of the incident electron is studied as a function of the
  magnetic flux (and vice versa) for both the periodic and quasiperiodic
  arrays of rings having different number of atoms in the arms. We find
  interesting resonance properties at specific values of the flux, as
  well as a power-law decay in the transmission coefficient as the
  number of rings increases, when the magnetic field is switched off.
  For the quasiperiodic Fibonacci sequence we discuss various features
  of the transmission characteristics as functions of energy and flux,
  including one special case where, at a special value of the energy and
  in the absence of any magnetic field, the transmittivity changes
  periodically as a function of the system size.
\end{abstract}

\pacs{
73.21.-b, 
73.23.-b, 
85.35.Ds 
}

\maketitle


\section{Introduction}
\label{sec-introduction}

Quantum transport in mesoscopic systems has been an exciting field
of research in the past several years
\cite{AltAS81,WebWUL85,ShaS81,GijVB84,PanCRG84,GefIY84,But86,ButIA84,SivI86,EntHF86,LeeSF90,LevDDB90,KowSEI90,HuM91,Xia92}.
One of the important aspects that has attracted much attention,
both experimentally and theoretically, is the fluctuation of the
magneto-conductance due to quantum coherence in such samples. For
mesoscopic systems at very low temperatures the phonon scattering
is insignificant, and the phase coherence length of the electrons
becomes large compared to the system size. In the presence of a
magnetic field one observes a specific, reproducible fluctuation
pattern of the conductance, as the magnetic field or the Fermi
level varies. The sample becomes equivalent to an electron
waveguide where the transport properties are determined by the
impurity configuration and the geometry of the conductor. In
experimental and theoretical works, the Aharonov-Bohm (AB) effects
\cite{AltAS81,WebWUL85} in solid state devices in the forms of
rings and cylinders that enclose a magnetic flux $\phi$ have been
investigated with much vigor
\cite{AltAS81,WebWUL85,ShaS81,GijVB84,PanCRG84,GefIY84,But86,ButIA84,SivI86,EntHF86,LeeSF90,LevDDB90,KowSEI90,HuM91,Xia92}.
The oscillations in the magneto-resistance were predicted to be
dominated by a half-integer flux period $\phi_0/2$ \cite{AltAS81},
where $\phi_0=hc/e$ is the fundamental flux quantum. This was
found experimentally for long cylinders
\cite{AltAS81,ShaS81,GijVB84} and arrays of metal rings
\cite{PanCRG84}. For single rings, it has been discussed that both
periods can be present \cite{WebWUL85}.  Among the theoretical
studies on single ring systems, D'Amato {\em et al.}\
\cite{DamPW89}, have discussed a tight-binding model of a
disordered ring coupled to two external leads, and have calculated
the Landauer conductance \cite{Lan70} as a function of $\phi$. For
strongly disordered rings and for arbitrary disorder with weakly
coupled branches, they have found a dominant period $\phi_0$.
Aldea {\em et
  al.}\ \cite{AldGC92} investigated the dc magneto-resistance in a
two-probe ring geometry within a tight-binding formalism. They
have presented analytical results for ordered single- and
double-ring structures and discussed the localization effects due
to disorder in such systems. AB effects for bound states such as
neutral excitonic particles have also recently been studied
\cite{WarSHB00,Cha95,RomR00}.

In comparison, the behavior of the magneto-conductance in systems having
serially connected rings, has received little attention. In one of the
early studies on serially connected rings Deo and Jayannavar
\cite{SinJ94,JayS95} discussed quantum transport in these systems. The
`band formation' in such geometries has been analyzed and some magnetic
properties of loop structures in the presence of an AB flux have been
discussed. Takai and Ohta \cite{TakO93} have also addressed similar
problems where both the magnetic flux and an electrostatic potential are
present. These works have relied on the solution of the continuous
version of the Schr\"{o}dinger equation for the ring systems as well as
other geometries \cite{SinJ94,JayS95,TakO93}.  Transmission through a
serial arrangement of rings can equivalently be handled within a
tight-binding formalism, in the spirit of the single-ring studies of
Ref.\ \cite{DamPW89} and Ref.\ \cite{AldGC92}.  However, only recently
an attempt has been made in this direction by Li {\em et al.}\
\cite{LiZL97}.  They used a tight-binding model and a scattering-matrix
technique to obtain closed expressions for transmission across a
periodic array of identical rings.

The tight-binding formalism naturally facilitates the application of
real-space renormalization-group methods to determine the bands and the
transmittivity. It is also easy to incorporate disorder in this scheme.
Additionally, it has also been successfully used to compute the dc
conductivity of quasi-one-dimensional polyaniline chains \cite{HeyS97},
which resemble the ring-like mesoscopic objects quite closely. In this
context, we believe that there is still scope to look deeper into the
transport properties of serially connected rings. For example, a
detailed analysis of the variation of the transmission properties as the
number of atoms (scatterers) in each arm of the ring changes, keeping
the electron energy $E$ and the flux $\phi$ constant, has been somewhat
less attended to so far.  Also, the effect of having rings arranged in
geometries other than periodic (quasiperiodic or random, for example) on
the overall transmittivity is something that has not drawn any attention
at all. This, to our mind, is worth investigating in detail.  In this
communication we address some of these problems.  We focus our attention
on the transmission coefficient across (i) perfectly ordered arrays of
identical rings, and (ii) across a quasiperiodic arrangement of the
rings of two different sizes.

Apart from the results that already exist and present an
interesting scenario, an additional motivation in undertaking the
present work comes from a recent experiment on the measurement of
low-temperature magnetic response of serially connected
GaAs/AlGaAs mesoscopic rings \cite{RabSMH01}. Persistent currents
with a period equal to $\phi_0$ have been detected, and it has
been pointed out that the persistent currents are not
significantly modified in connected systems. To our mind, this
sort of an experiment generates some interest in studying also the
transport if rings are arranged in a quasiperiodic sequence.

We confine ourselves to rings where all the sites have identical
`on-site potential', and are connected to the nearest neighbors by
a hopping integral of the same magnitude.  We provide a formula to
reduce this ring geometry to an effective `dimer'. An array of
rings then becomes equal to an array of such dimers, where the
essence of quasi-one dimensionality is taken care of in an exact
way. For a sequence of identical rings (identical dimers) our
method allows us to analyze the variation of the transmittivity as
a function of the magnetic flux at a fixed value of the electron
energy, and vice versa.  The situation, when both the energy and
the flux vary, can be dealt with easily in this formalism. We show
that the role of the number of scatterers contained in each arm of
the ring is significant in determining the transmission across a
series of rings, and sometimes may lead to rather unexpected
behavior. For example, we show analytically that, when the two
arms of a ring contain even and odd number of atoms respectively,
and the flux threading the rings is zero, the transmission
coefficient for $N$ such rings exhibits a {\it power-law decay},
viz.\ $T \sim 1/N^2$ for large $N$, if we choose the electron
energy $E=\epsilon$, $\epsilon$ being the on-site potential
corresponding to an atom in the ring. The power-law decay however,
is sensitive to the choice of the parameters in the leads which
are to be connected to the two ends of the system under study.
This aspect will be discussed in more detail below.

We further investigate the electronic properties of quasiperiodically
arranged sequences of rings. In particular, we study the spectral
features of a Fibonacci arrangement of rings of two different sizes in
the presence of the same magnetic flux per area through each ring. We
find a non-trivial modification of the energy `bands' for such a
multiple-ring system. A quasiperiodic sequence of rings may even display
a periodic variation of the transmittivity as a function of the system
size in the absence of a magnetic field at specific values of the
electron energy.  We discuss one such case in detail.

In section \ref{sec-method} we present the basic method of our
calculation for the transmission coefficient, and apply it to
discuss some of the results for the isolated rings. The periodic
arrangement of the rings is considered in section
\ref{sec-periodic}.  In section \ref{sec-quasiperiodic} we focus
on the quasiperiodic geometries. The transmittivity for a
Fibonacci array of rings at an arbitrary generation of the series
has been calculated by suitably modifying the trace-antitrace
formulation \cite{WanGS00} to include the magnetic flux. In
section \ref{sec-conclusions} we summarize.

\section{From a ring to a dimer}
\label{sec-method}

\begin{figure}
  \centering
  \includegraphics[width=0.6\columnwidth]{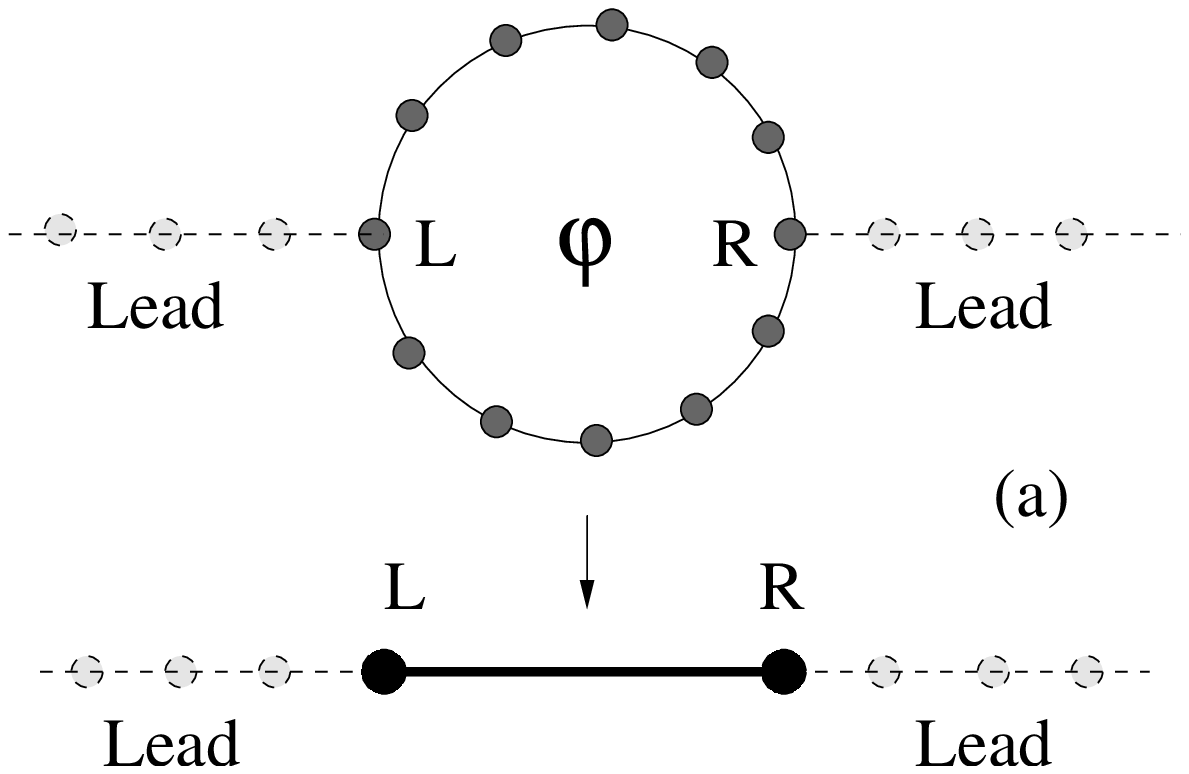}\\
  \vspace*{5ex}
  \includegraphics[width=0.95\columnwidth]{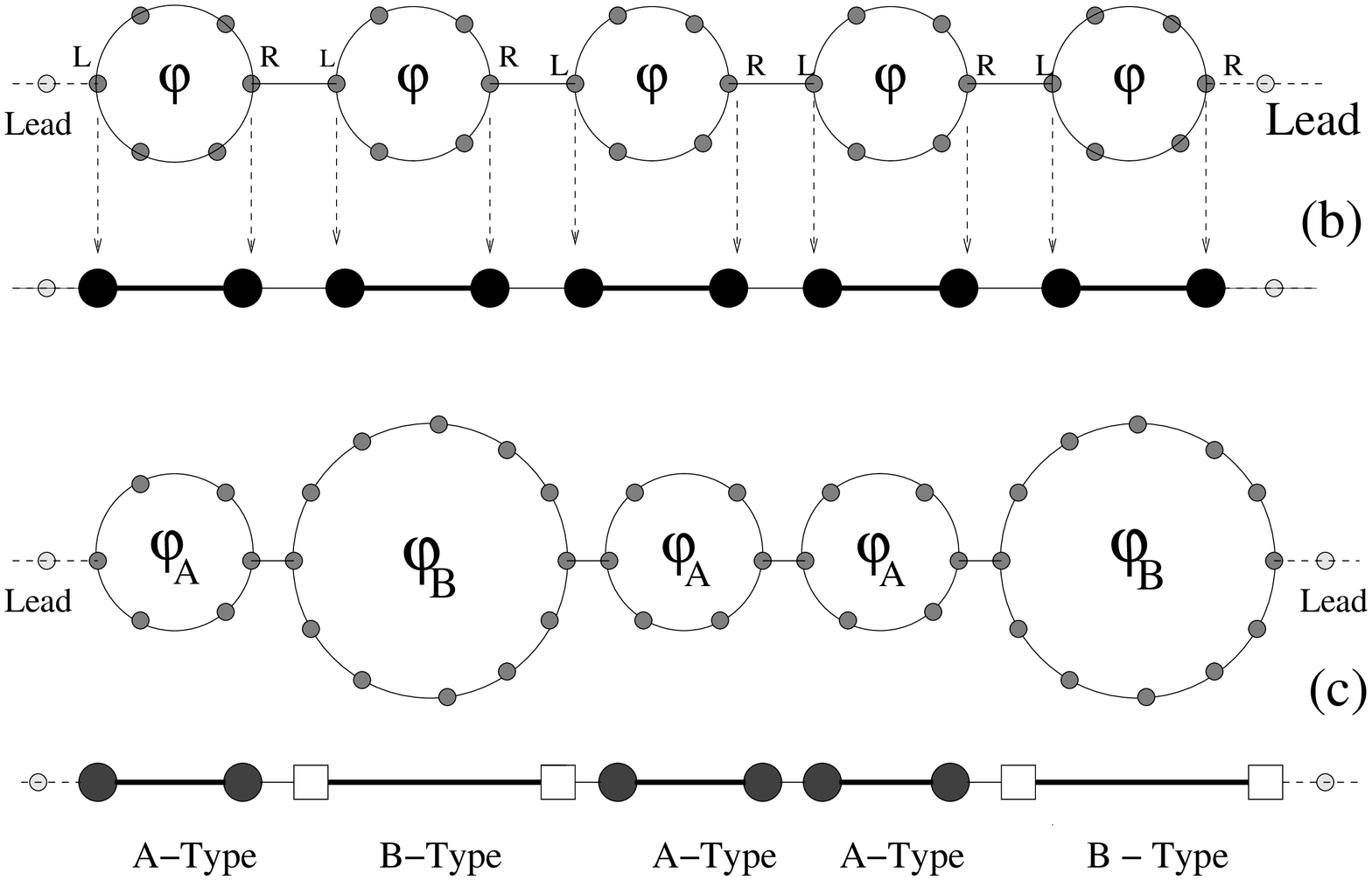}
  \caption{\label{fig1}
    Rings in the tight-binding model and the effective dimers.  (a)
    Single ring getting reduced to a dimer.  (b) An ordered array of
    identical rings and the array of identical dimers which follow.  (c)
    A Fibonacci array of rings of two different sizes and the
    corresponding dimer structure. In all cases, a thin solid line
    denotes the `bulk' hopping amplitude $t$, a thick solid line
    corresponds to a renormalized dimer hopping $\tilde{t}_{F,B}$ and a
    dashed line represents the hopping $t_0$ in the leads.}
\end{figure}

In this section we explain how we reduce a single ring geometry to a
dimer. Let us concentrate on the simplest model of a ring which
consists of identical `atoms' each characterized by the same
on-site potential energy $\epsilon$. The atoms are assumed to be
equally spaced on the ring.  The ring is attached at two sites $L$
and $R$ to two semi-infinite ordered leads to facilitate the
transmission measurement as shown in Fig.\ \ref{fig1}(a).
The leads are characterized by a constant on-site potential $\epsilon_0$
and a uniform hopping integral $t_0$ between the nearest neighbour sites.
The leads can be attached to any two atoms, which is equivalent to
saying that we may have different number of atoms in the `upper' and
`lower' arms of the rings. A magnetic field of flux density $\phi$
penetrates the ring. We describe such a ring by the standard
tight-binding Hamiltonian with the Peierls' substitution \cite{Pei33},
\begin{equation}
  H = \epsilon\sum_i|i\rangle\langle i| + t\sum_{ij}\left[e^{i\gamma}
    |i\rangle \langle j|+e^{-i\gamma}|j\rangle\langle i|\right]
\end {equation}
where $t$ is the amplitude of the hopping integral for
nearest-neighbor couplings of identical strength, and $|i\rangle$,
$i=1, \ldots, N_0$ denotes the tight-binding orbitals. The flux
$\phi$ is measured in units of $\phi_0$ and enters the Hamiltonian
via $\gamma=2\pi\phi/N_0\phi_0$. $N_0$ is the number of bonds (and
sites) in the ring.  The amplitudes of the wave function at the
$j$th site in any arm of the ring (excluding the sites $L$ and $R$
at the junctions with the leads) are related to the amplitudes at
the nearest-neighbor sites by the difference equation,
\begin{equation}
(E-\epsilon)\psi_j = t e^{i\gamma}\psi_{j+1} + t e^{-i\gamma}\psi_{j-1}.
\end{equation}
In order to calculate the transmission across this structure,
first we `renormalize' the ring into a dimer \cite{DamPW89}
comprising just two `modified' atoms (cp.\ Fig.\ \ref{fig1}(a)),
with on-site potential $\tilde\epsilon$ connected by an effective
hopping integral with amplitude $\tilde t$. If the two arms of the
ring contain an unequal number of atoms, the effective hopping
integral is complex, i.e., $\tilde t$ contains a phase that
reflects the broken time-reversal symmetry between the components
of the dimer. To obtain this, we need to eliminate all the atoms
that lie between the two points of contact $L$ and $R$ as shown in
Fig.\ \ref{fig1}(a). This can been done analytically. It is
straightforward to relate the amplitudes $(\psi_j,\psi_{j-1})$ on
any arm of the ring to the set $(\psi_{j-1},\psi_{j-2})$ on the
same arm through a $2\times 2$ transfer matrix
\begin{equation}
\mbox{\boldmath $M$}=\left( \begin{array}{cc}
\frac{E-\epsilon}{t}e^{-i\gamma}&-e^{-2i\gamma}\\1&0
\end{array}\right) .
\label{eq-tm22}
\end{equation}
The matrix has a determinant equal to $\exp(-2i\gamma)$.  The product
of $n$ such matrices is
\begin{eqnarray}
  \mbox{\boldmath $M$}^n &=&
  e^{-i(n-1)\gamma}\left[U_{n-1}\left(\frac{E-\epsilon}{2t}\right)\mbox{\boldmath
      $M$} \right.\nonumber \\
      & & \mbox{ }
      \left. -e^{-i\gamma}U_{n-2}\left(\frac{E-\epsilon}{2t}\right)\openone\right]
\end{eqnarray}
where $\openone$ is the $2\times 2$ identity matrix and $U_n$ is
the $n$th order Chebyshev polynomial of the second kind. Using
this result together with the set of difference equations $(2)$ we
obtain the expressions for the effective on-site potentials of the
contact points $L$ and $R$. For the general case of $n$
and $m$ atoms between $L$ and $R$ in the upper and
the lower arm (excluding the contact points $L$ and $R$),
 respectively, we get
\begin{equation}
  \tilde\epsilon = \epsilon - t e^{i\gamma}
\frac{M_{12}^{n}}{M_{11}^{n}}
  - t e^{-i\gamma}\frac{\left(M_{12}^{m}\right)^*}%
  {\left(M_{11}^{m}\right)^*}
\label{eq-effepsilon}
\end{equation}
where $\mbox{$M$}^{n}_{ij}$ is the $(i,j)$th element of
$\mbox{\boldmath $M$}^{n}$. The effective hopping
integral includes the effect of the broken time-reversal symmetry
resulting from the application of the magnetic field, and is
denoted as $\tilde t_F$ and $\tilde t_B$ corresponding to the
`forward' and the `backward' hopping. The forward hopping integral
is given by
\begin{equation}
  \tilde t_F = \frac{t e^{i\gamma}}{ \mbox{$M$}_{11}^{n}}
+ \frac{t e^{-i\gamma}}{ \left(M_{11}^{m}\right)^*}
\label{eq-effthop}
\end{equation}
and $\tilde t_B=\tilde t_F^*$. The ring embedded in the ordered lead
now reduces to a dimer comprising of the (modified) sites $L$ and
$R$.  The transfer matrices for these two sites are
\begin{equation}
\label{eq-ML}
\mbox{\boldmath $M$}(L)=\left( \begin{array}{cc}
\frac{E-\tilde{\epsilon}}{\tilde t_F}&-\frac{t_0}{\tilde t_F}\\1&0
\end{array}\right)
\end{equation}
and
\begin{equation}
\label{eq-MR}
\mbox{\boldmath $M$}(R)=\left( \begin{array}{cc}
\frac{E-\tilde{\epsilon}}{t_0}&-\frac{\tilde t_B}{t_0}\\1&0
\end{array}\right)
\end{equation}
with $t_0$ the hopping integral in the lead. The next step is to
calculate the product {\boldmath $P$} of the two matrices
corresponding to the two sites $R$ and $L$, i.e., $\mbox{\boldmath
$P$}= \mbox{\boldmath $M$}(R)\mbox{\boldmath
  $M$}(L)$. Using a well known formula \cite{StoJC81} the transmission
coefficient of this effective dimer $L$-$R$ can then be obtained
as
\begin{widetext}
\begin{equation}
  T = \frac{4 \sin^2 k}{|\mbox{$P$}_{12}-
\mbox{$P$}_{21}+(\mbox{$P$}_{11}-\mbox{$P$}_{22})\cos k|^2 +
    |\mbox{$P$}_{11}+\mbox{$P$}_{22}|^2 \sin^2 k}
\label{eq-transmission}
\end{equation}
\end{widetext}
with parametrization $E=\epsilon_0 + 2t_0 \cos k$. The lattice
spacing has been chosen to be unity throughout.  The above scheme
has been tested to reproduce the results of the single-ring cases
\cite{AldGC92} accurately.

We emphasize that the choice of the lead parameters is of much
importance. For a given set of values of $\epsilon_0$ and $t_0$ in
the lead we will be able to scan only those energy eigenvalues of
our system which fall within the allowed `band' of the lead, i.e.,
from $\epsilon_0-2t_0$ to $\epsilon_0+2t_0$. Thus it is in
principle possible to choose ring parameters $\epsilon$ and $t$
such that there will be energies allowed in the ring structures
that will be masked by the leads. Nevertheless, in the
experimental situation the leads (not to be mistaken with the {\em
contacts}) are usually made of the same material as the ring
system. Therefore, we restrict ourselves to uniform values $t_0=t$
and $\epsilon_0=\epsilon$ throughout the system and therefore to
an accessible energy band $\epsilon-2t$ to $\epsilon+2t$  in the
following.

\section{Periodic array of rings}
\label{sec-periodic}

In this section we consider the variation of the transmission
coefficient as a function of the flux at some fixed value of the
energy.  For convenience we restrict ourselves to the energy
$E=\epsilon$, for which the Chebyshev polynomials assume particularly
simple forms, and precise analytical expressions are obtained for
the effective on-site potential $\tilde\epsilon$ and the hopping
term $\tilde t_F$ and its complex conjugate. The
hopping integral connecting one ring to the next is assumed to be same as
that between the atoms in the rings, i.e., equal to $t$.

\subsection{Even-even case:}

\begin{figure}
  \figspace
  \centerline{\includegraphics[width=0.95\columnwidth]{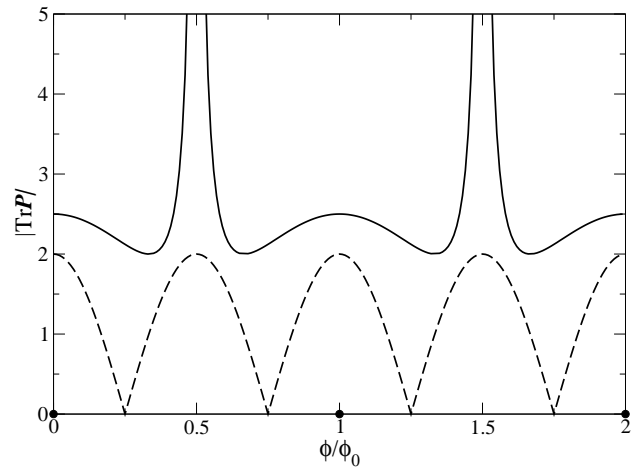}}
  \caption{\label{fig2}
    Variation of ${\rm Tr}\mbox{\boldmath $P$}$ against flux
    $\phi/\phi_0$ for $E=\epsilon=0$. Here, $p=2$ and $t=1$. The solid
    and dashed lines correspond to, respectively, the even-even and the
    even-odd cases discussed in section \protect\ref{sec-periodic}. For
    the odd-odd case the trace diverges except at $\phi=0$, $\phi=\phi_0$
    and $\phi=2 \phi_0$ where is takes on the value zero as shown by solid
    dots at these points.}
\end{figure}

\begin{figure}
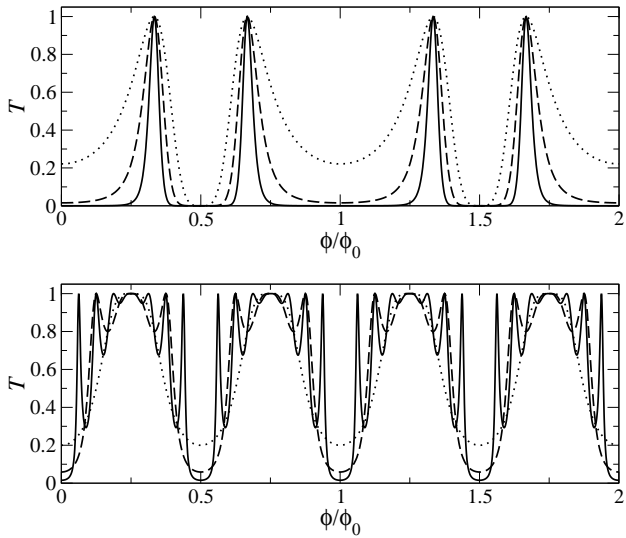

  \centering \figspace
  \centerline{\includegraphics[width=0.95\columnwidth]{figs/fig-T248_phi_EE.eps}}
  \figspace
  \centerline{\includegraphics[width=0.95\columnwidth]{figs/fig-T248_phi_EO.eps}}

  \caption{\label{fig3}
    (a) Transmission coefficient $T$ versus magnetic flux $\phi$ for an
    array of identical rings with $n=m=4$,
    (b) for $n=5$ and $m=4$.
%
    The dotted, dashed, and solid lines correspond
    to two-, four- and eight-ring systems, respectively.  We display
    flux values up to $\phi=2\phi_0$ in order to get a view of the
    periodicity.  We have set $\epsilon=0$ and $t=1$ for the atoms on
    the rings, and $\epsilon_0=0$ and $t_0=1$ for those in the lead
    throughout. }
\end{figure}

\begin{figure}[ht]
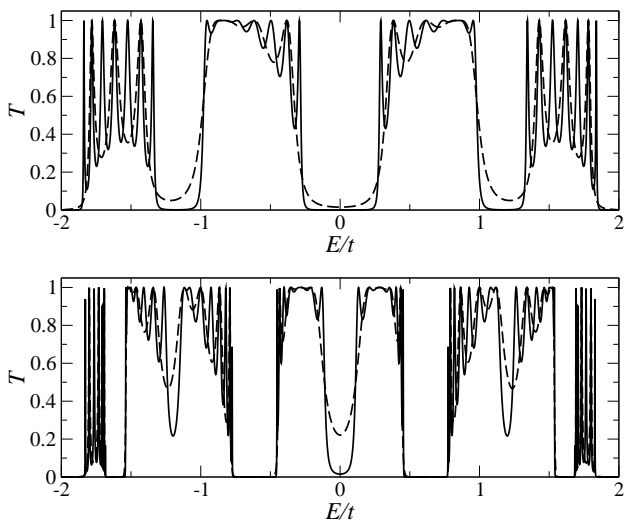

  \figspace
  \centerline{\includegraphics[width=0.95\columnwidth]{figs/fig-T248-phi0_E.eps}}
  \figspace
  \centerline{\includegraphics[width=0.95\columnwidth]{figs/fig-T248-phi4_E.eps}}

  \caption{\label{fig3-b}
    (a) Transmission coefficient $T$ versus energy $E$ for the same
    system as in Fig.\ \protect\ref{fig3} with (a) $\phi=0$.
    (b) $T$-$E$ diagram for $\phi=\phi_0/4$.  The dashed
    and the solid lines correspond to four- and eight-ring
    systems, respectively.}
\end{figure}

We take $n=m=2p$ with $p$ an integer. The
number of sites in the ring, including the contact points, is
$4p+2$. Using the values of the appropriate Chebyshev polynomials
at $E = \epsilon$, it is not difficult to work out with
the help of Eqs.\ (\ref{eq-effepsilon}) and (\ref{eq-effthop})
that $\tilde{\epsilon} = 0$ and $\tilde t_F = \tilde t_B = (-1)^p
2t \cos(\pi\phi/\phi_0)$.  The time-reversal symmetry is preserved
in this case.
Let us check if $E=\epsilon$ belongs to the spectrum of an infinite array
of such rings.
At $E=\epsilon$ the product transfer matrix
corresponding to the dimer, which constitutes a `unit cell' of
the infinite array, becomes
\begin{equation}
\label{eq-Pee}
\mbox{\boldmath $P$}_{{\rm even-even}}=
(-1)^{p+1} \left( \begin{array}{cc}
\frac{\sigma}{t} & 0 \\
0 & \frac{t}{\sigma}
\end{array}\right)
\end{equation}
where $\sigma = 2t \cos(\pi\phi/\phi_0)$.
In order to have
$E=\epsilon$ in the spectrum of an infinite ordered array of such
rings, one must have $|{\rm Tr}\mbox{\boldmath $P$}|\le 2$.
However, from (\ref{eq-Pee}) it is evident that $|{\rm
Tr}\mbox{\boldmath $P$}|=2$ only for $\phi/\phi_0=1/3, 2/3, 4/3,
5/3, \ldots$. At all other flux values the trace is greater than
$2$ as shown in Fig.\ \ref{fig2}.
Therefore, at $E=\epsilon$ and at arbitrary flux, not equal to a
special value such as above, the transmission coefficient across
an array of this type of rings will decay exponentially as the
system increases in size. This can also be checked using the
formula for the transmission coefficient. For the special values
of the flux mentioned above the transmission coefficient is
precisely {\it unity}. This implies that for $E=\epsilon$, we can
achieve totally ballistic transport by tuning the flux to a
specific value.  This is an example for an extended eigenstate
under the influence of a magnetic field
\cite{AldGC92,SinJ94,JayS95}. The phenomenon of full transmission
at these specific flux values can be understood if we look at the
values of $\tilde\epsilon$ and $\tilde t_F=\tilde t_B$. We have
presented results here for $\epsilon= \epsilon_0 = 0$. Now,
$\tilde\epsilon$ is zero, and equal to the on-site terms at the
leads. As $\tilde t_F$ also becomes real and equal to unity, the
electron essentially `sees' an ordered array of identical sites
connected by identical hopping integrals. The corresponding
eigenstate is naturally extended and $E=0$ happens to be the band
center. In Fig.\ \ref{fig3}(a) we display the variation of the
transmission coefficient against flux for this case.
The transmission coefficient is periodic in flux with a period
equal to $\phi_0$. As the number of rings increases the increase
in the sharpness of transmission at specific flux values is
evident. In Fig.\ \ref{fig3-b}, we display the variation of the
transmission coefficient as a function of the energy $E$.
The anti-resonance ($T=0$) sets in clearly for the eight-ring
series when $\phi=0$. The overall transmission for $\phi=\phi_0/4$
appears to be slightly enhanced compared to $\phi=0$. If we select
$\phi=\phi_0/2$, an array of arbitrary size becomes completely
opaque to an incoming electron with $E=0$. This is also clear from
(\ref{eq-Pee}) and consistent with the findings in the single ring
case \cite{AldGC92}.

\subsection{Odd-odd case:}

We now consider an equal but odd number of atoms in each arm,
i.e., $n=m=2p+1$ such that the total number of atoms in the ring
remains even. In this case $\mbox{\boldmath $M$}_{11}^n = 0$ at
$E=\epsilon$. Both $\tilde \epsilon$ and $\tilde t$ diverge as $E
\rightarrow \epsilon$, leading to a divergence of the trace,
except at some special values of the flux for which the trace
becomes exactly equal to zero leading to perfect transmission. An
analytical attempt can be made to see this in the following way.
Using Eqs.\ (\ref{eq-effepsilon}) and (\ref{eq-effthop}) one can
show that, in the limit $E \rightarrow \epsilon$, $\tilde \epsilon
= \epsilon + \frac{2t^2}{(p+1)(E-\epsilon)}$ and $\tilde t_F =
(-1)^p \frac{t\sigma}{(p+1)(E-\epsilon)}$, where the leading terms
in the expressions for $\tilde{\epsilon}$ and $\tilde{t}_F$ have
been retained as $E \rightarrow \epsilon$. Here also $\tilde
t_{B}=\tilde {t}_F$. The elements of the transfer matrix across
the dimer turn out to be
\begin{subequations}
\begin{eqnarray}
  \label{eq-Poddodd}
  P_{\rm odd-odd}^{11} & = & (-1)^p \left\{
\left[\frac{(p+1)(E-\epsilon)^2-2t^2}
{\sigma t}\right]^2 - 1 \right\} \nonumber \\
& & \mbox{ } \times
\frac{\sigma}{(p+1)(E-\epsilon)} \\
P_{\rm odd-odd}^{12} & = & (-1)^{p+1}
\frac{(p+1)(E-\epsilon)^2 - 2t^2}{t\sigma} \\
P_{\rm odd-odd}^{21} & = & - P_{\rm odd-odd}^{12} \\
P_{\rm odd-odd}^{22} & = & (-1)^{p+1}
\frac{(p+1)(E-\epsilon)}{\sigma} \quad .
\end{eqnarray}
\end{subequations}
It can be easily verified that $|{\rm Tr}\mbox{\boldmath $P$}_{\rm
  odd-odd}|=0$ for $\phi = m \phi_0$, with $m=0, 1, 2, \ldots$.
  For all other values the trace diverges as $E
\rightarrow \epsilon$. Thus the transmission coefficient across an
arbitrarily long array of the above rings is unity at the
specified values of the flux and zero otherwise. This can easily
be worked out using Eq.\ (\ref{eq-transmission}). There is no
dependence of the transmission coefficient on the size of the
system.  For the zero flux case, the $T$-$E$ diagram exhibits
resonance at $E=0$, in contrast to the previous case.

\subsection{Even-odd case:}

We next take $n=2p$ and $m=2p+1$. Proceeding in the same spirit as
in the odd-odd case, the effective on-site potential and the
hopping matrix elements of the dimer read $\tilde \epsilon =
\epsilon + \frac{t^2}{(p+1)(E-\epsilon)}$ and, $\tilde t_F =
(-1)^p \frac{t\beta}{(p+1)(E-\epsilon)}$, where the leading terms
in the expressions for $\tilde \epsilon$ and $\tilde t_F$ have
been retained as $E \rightarrow \epsilon$. Here,
$\beta=(p+1)(E-\epsilon) \exp[i(2p+1)\chi]+ t \exp[-i(2p+2)\chi]$,
and $\chi=\frac{2\pi\phi/\phi_0}{4p+3}$. The elements of
$\mbox{\boldmath $P$}_{\rm
  even-odd}$ are
\begin{subequations}
\begin{eqnarray}
  \mbox{$P$}_{\rm even-odd}^{11} & = & (-1)^p \left\{ \left[
      \frac{(p+1)(E-\epsilon)^2-t^2}{t\beta}\right]^2 \beta
        -\beta^{*}\right\}
\nonumber \\
& & \mbox{ } \times
  \frac{1}{(p+1)(E-\epsilon)} \\
  \mbox{$P$}_{\rm even-odd}^{12} & = & (-1)^{p+1}
\frac{(p+1)(E-\epsilon)^2-t^2}{t\beta}
  \\
  \mbox{$P$}_{\rm even-odd}^{21} & = & -\mbox{$P$}_{\rm even-odd}^{12} \\
  \mbox{$P$}_{\rm even-odd}^{22} & = & (-1)^{p+1}
    \frac{(p+1)(E-\epsilon)}{\beta}
  \quad .
\end{eqnarray}
\end{subequations}
It is not difficult to work out analytically that, as $E
\rightarrow \epsilon$, the maximum value of $|{\rm Tr}
\mbox{\boldmath $P$}_{\rm even-odd}|$ is $2$. That is, the trace
is always bounded by $2$ from above.  This implies that
$E=\epsilon$ is in the spectrum of an infinite array of these
rings for all values of the flux. In Fig.\ \ref{fig2} we show the
variation of $|{\rm Tr} \mbox{\boldmath $P$}_{\rm even-odd}|$
against $\phi/\phi_0$.

To study the transmission coefficient, let us first consider the
case $\phi=0$ such that
\begin{equation}
\label{eq-10} \mbox{\boldmath $P$}_{\rm even-odd}({\phi=0})=
(-1)^p \left( \begin{array}{cc} -2&1\\-1&0
\end{array}\right) .
\end{equation}
A product of such matrices will look like \cite{JordanForm}
\begin{equation}
\mbox{\boldmath $P$}_{\rm even-odd}^N({\phi=0}) = \left(
\begin{array}{cc} N+1&-N\\N&1-N
\end{array}\right) .
\end{equation}
when $N$ is even. A similar expression can be worked out for odd
values of $N$.  In either case, it can be checked from Eq.\
(\ref{eq-transmission}) that with $\epsilon_0=\epsilon=0$ and
$t_0=t=1$, the transmission coefficient exhibits a {\it power-law
decay} for large values of $N$, i.e., $T \sim 1/N^2$. At $E=0$,
the matrix \mbox{\boldmath $P$}$_{\rm even-odd}$ is exactly the
same as the transfer matrix for a periodic chain having $N$ sites
with $\epsilon=0$ with the electron at energy $E=-2t$ (which
defines the band edge of the infinite ordered chain).  Hence the
($E=0,\phi=0$)-combination for a periodic array of even-odd rings
is equivalent to the $E=-2t$ situation of a periodic chain of
atoms. Just as we analyzed the even-even case, here it is
straightforward to show that unit transmission can be achieved by
tuning the magnetic field so that $\phi=(2m+1)\phi_0/4$,
irrespective of the size of the system.
We show the variation of the transmission coefficient against
$\phi/\phi_0$ in Fig.\ \ref{fig3}(b).  The transmission
coefficient is periodic, with a period equal to $\phi_0/2$.

\section{Quasiperiodic array of rings}
\label{sec-quasiperiodic}

\subsection{Fibonacci array}

We follow the usual method of building a Fibonacci sequence
recursively \cite{KohKT83}. The first objective will be to study
the effect of the sequence of rings shown in Fig.\ \ref{fig1} (c)
on the `band structure' of the system as it grows in size. We
model the system by placing two different rings in series
following the growth rule $A \rightarrow AB$ and $B \rightarrow
A$. Here, $A$ and $B$ stand for two rings of different sizes, but
having identical on-site potentials and hopping integrals (Fig.\
\ref{fig1}(c)).  This, to our mind, represents a model that can
possibly be realized experimentally \cite{RabSMH01}.
We assume that the system is immersed in a constant magnetic
field so that the flux $\phi_A$ and $\phi_B$ through the rings is
proportional to their respective areas $S_A$ and $S_B$. $\gamma$
as defined in section \ref{sec-method} now takes on two different
values, $\gamma_A$ and $\gamma_B$ related via
\begin{equation}
\gamma_B = \frac{N_B}{N_A} \gamma_A
\end{equation}
with $N_A$ and $N_B$ denoting the total number of scatterers in $A$-type and
$B$-type rings, respectively.

Following the procedure described in section \ref{sec-method}, we
reduce each ring to an effective dimer, so that we finally have
a Fibonacci sequence of these dimer-like objects $A$ and $B$,
characterized by $\mbox{\boldmath $P$}_A$ and $\mbox{\boldmath
$P$}_B$, respectively, which can be easily computed from
(\ref{eq-ML}), (\ref{eq-MR}). The product transfer matrix across
the $m$th generation Fibonacci chain is, as usual \cite{KohKT83},
\begin{equation}
  \mbox{\boldmath $M$}(m+1) = \mbox{\boldmath $M$}(m-1)\mbox{\boldmath
    $M$}(m)
\end{equation}
with $\mbox{\boldmath $M$}(1)=\mbox{\boldmath $P$}_A$ and,
$\mbox{\boldmath $M$}(2)=\mbox{\boldmath $P$}_B \mbox{\boldmath
  $P$}_A$.  Keeping in mind that these matrices have a phase factor
similar to $\exp(-i\gamma)$ as in (\ref{eq-tm22}), we have generalized
the trace map \cite{KohKT83} relation to
\begin{equation}
x_{m+1} = x_m x_{m-1} - \Delta_{m-1} x_{m-2}
\label{eq-gentracemap}
\end{equation}
where $x_m={\rm Tr}\mbox{\boldmath $M$}(m)$, and not the
`half-trace' as usual \cite{KohKT83}, and $\Delta_m= \det
\mbox{\boldmath $M$}(m)$.  At the start of the sequence, we have
$x_1={\rm  Tr}\mbox{\boldmath $P$}_A$ and $x_2={\rm
Tr}(\mbox{\boldmath $P$}_B \mbox{\boldmath $P$}_A)$.
The above formula enables us to deal with situations where the
phase accumulated by the electron in traversing one arm of the
ring is {\it not canceled} by the phase accumulated from a trip
along the other arm.

\begin{figure}

  \figspace
  \centerline{\includegraphics[width=0.95\columnwidth]{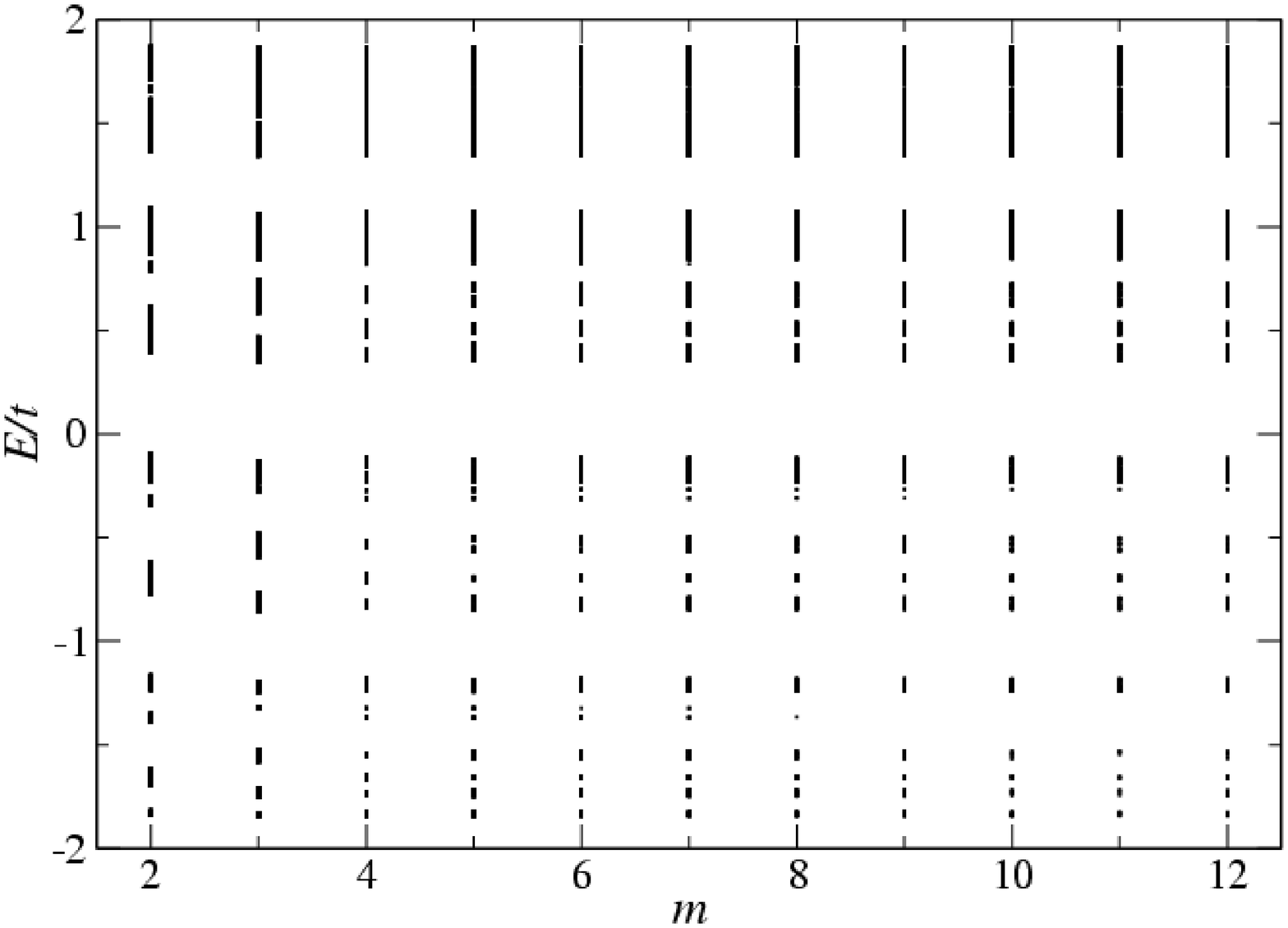}}
  \figspace
  \centerline{\includegraphics[width=0.95\columnwidth]{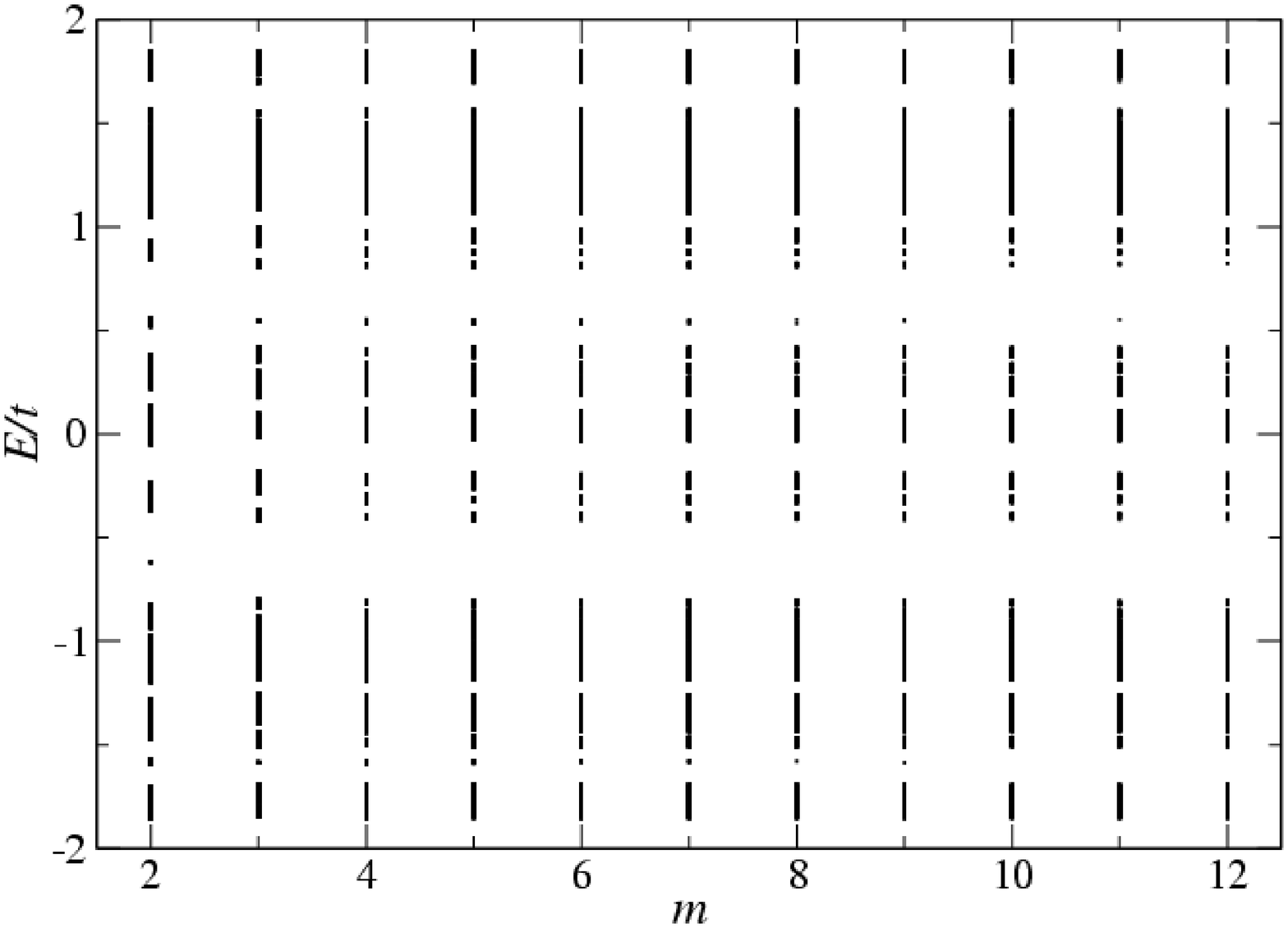}}

  \caption{\label{fig4}
    Distribution of the energy eigenvalues for a Fibonacci array of
    rings plotted against the generation index $m$: (a) with $\phi_A=0$
    and (b) with $\phi_A=\phi_0/4$. The $A$-ring has $4$ atoms in each
    arm, whereas the $B$-ring has $5$ atoms in the upper arm and $4$ in
    the lower arm. The energy range has been scanned at an interval of
    $\Delta E=0.0001 t$.}
\end{figure}

In Fig.\ \ref{fig4} we show the distribution of the allowed energy
values which correspond to $|x_m| \le 2$ \cite{KohKT83} within an
energy range $[-2t,2t]$.
The use of the above trace-formula is {\it essential} in this
case. We have also studied the change in the character of the
`band' as a magnetic field is switched on. Fig.\ \ref{fig4}(b)
shows the allowed energy values for $\phi_A=\phi_0/4$
($\phi_B\equiv\frac{N_B^2}{N_A^2}\phi_A=\frac{11^2}{10^2}
\phi_A$).  The pattern shows a marked change with respect to the
earlier case.

\begin{figure}
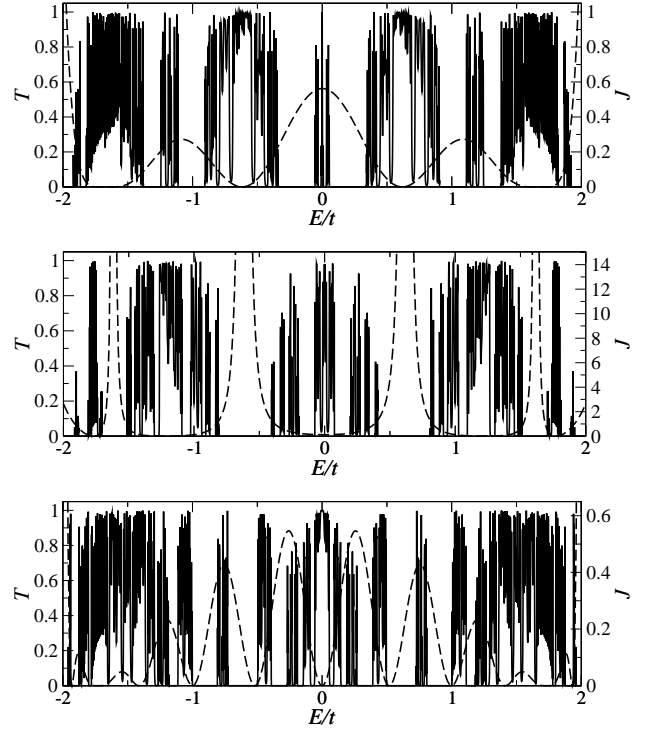


  \figspace
  \centerline{\includegraphics[width=0.95\columnwidth]{figs/fig-TFiBo9-phi0_E.eps}}
  \figspace
  \centerline{\includegraphics[width=0.95\columnwidth]{figs/fig-TFiBo9-phi4_E.eps}}
  \figspace
  \centerline{\includegraphics[width=0.95\columnwidth]{figs/fig-TFiBo9b-phi0_E.eps}}

  \caption{\label{fig5}
    Variation of the transmission coefficient $T$ against $E$ for a 9th
    generation Fibonacci array of rings (solid line) and the invariant
    $J$ (dashed line) for (a) $\phi_A=0$ and (b) $\phi_A=\phi_0/4$.
    $n=m=4$ for the $A$-type ring and $n=m=9$ for the $B$-type ring.
    (c) Same as (a), but $n=m=3$ for $A$ and $n=m=15$ for $B$. The
    energy resolution in all cases is $\Delta E/t= 0.001$.}
\end{figure}

It is also interesting to note that for such a quasiperiodic Fibonacci
array of rings the map (\ref{eq-gentracemap}) leads to an invariant in
the conventional  sense \cite{KohKT83}. The invariant in this case is a
function of $E$, $\phi_A$ and $\phi_B$, and is given by
\begin{equation}
J  =  \frac{1}{4}\left( \frac{x_{m+1}^2}{\Delta_{m+1}} +
    \frac{x_m^2}{\Delta_m} +
    \frac{x_{m-1}^2}{\Delta_{m-1}}
- \frac{x_{m+1} x_m
x_{m-1}}{\Delta_{m}\Delta_{m-1}} - 1 \right).
\end{equation}
For an ordered array of rings it is equal to zero. However, even
for an ordered array of rings we have resonance ($T=1$) and
anti-resonance ($T=0$) as a result of interference. Hence, the
zeros of the invariant $J$ should not necessarily correspond to
the $T=1$ cases in a Fibonacci array as well. In order to compute
the transmission coefficient for rings in a Fibonacci sequence we
generalize the trace-antitrace formulation discussed in the
literature \cite{WanGS00}. The transmission coefficient for an
$m$th generation sequence is given by \cite{WanGS00}
\begin{equation}
\label{eq-FiBo-Tm}
T(m) = \frac{4 \sin^2 k}{|z_m \cos k-y_m|^2 +
|x_m^2|\sin^2 k}
\end{equation}
where the antitraces $y_m$ and $z_m$ are given by
$y_m=\mbox{$M$}_{21}(m)- \mbox{$M$}_{12}(m)$ and
$z_m=\mbox{$M$}_{11}(m)-\mbox{$M$}_{22}(m)$.  They are obtained
recursively as
\begin{eqnarray}
y_{m+1} & = & x_m y_{m-1} + \frac{\Delta_m}{\Delta_{m-2}} y_{m-2} , \nonumber \\
z_{m+1} & = & x_m z_{m-1} + \frac{\Delta_m}{\Delta_{m-2}} z_{m-2} .
\end{eqnarray}
We emphasize that the use of Eq.\ (\ref{eq-FiBo-Tm}) in obtaining
$T(m)$ depends crucially on setting $t_0=t$ which, of course, has been
our choice here.
In Fig.\ \ref{fig5} we show the variation of transmission
coefficient and the invariant for a $9$th generation sequence with
$\phi_A=\phi_B=0$ and $\phi_A=\phi_0/4$, $\phi_B=4\phi_A$.
The $A$- and $B$-rings have $4$ and $9$ atoms in each arm,
respectively.  This implies that the circumference of the $B$-ring
is double the size of the $A$-ring.  The invariant is real in this
case. The zero-flux case is characterized by the appearance of
several subbands symmetrically distributed around $E=0$, while
with $\phi_A=\phi_0/4$, the subbands get closer. We have scanned
the energy range $[-2t,2t]$ using various values of the energy
interval $\Delta E$. The results are presented for relatively
large $\Delta E$ for more clarity in the figures.  On reducing the
energy interval further we find that the plot becomes more dense
within a subcluster. However, no new subband structure emerges for
the cases we present. With increasing number of scatterers the
invariant exhibits many more zeroes and the plot of the
transmission coefficient becomes highly fragmented, see Fig.\
\ref{fig5}(c). A similar effect is also observed for the periodic
arrays as well. It may be noted that the invariant becomes very
close to zero at certain points which correspond to high
transmittivity, and the ranges of the energy where the invariant
diverges correspond to the gaps in the spectrum. However, as we
have already mentioned, finite transmission may also be seen for
energies for which the invariant exhibits a finite value. Last,
the behavior of the transmission coefficient as shown above is by
no means unique and depends on the choice of the number of
scatterers in each ring. For example, if the $A$-ring has $3$
atoms in each arm, and the $B$-ring has $7$ atoms in each arm, the
size of the $B$-ring is again double the size of the $A$-ring.
However, in this case, for $\phi_A=\phi_B=0$ we have a
transmission maximum at $E=0$, in contrast to the case shown in
Fig.\ \ref{fig5}.

\subsection{Variation of $T$ against flux}

\begin{figure}
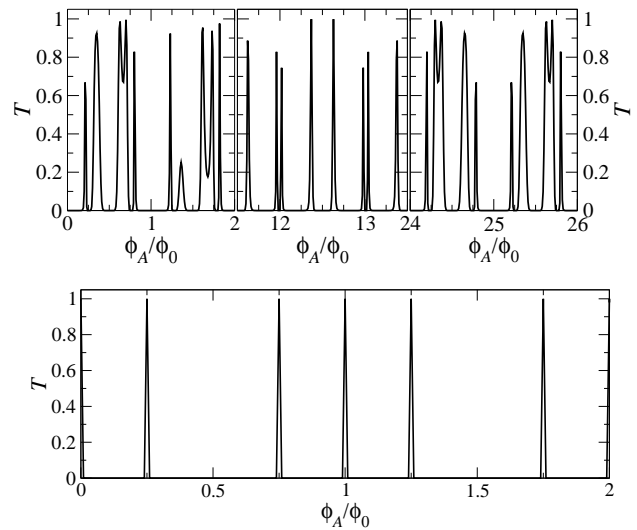


  \figspace
  \centerline{\includegraphics[width=0.95\columnwidth]{figs/fig-TFiBo9_phi_a.eps}}
  \figspace
  \centerline{\includegraphics[width=0.9\columnwidth]{figs/fig-TFiBo9_phi_b.eps}}

  \caption{\label{fig6}
    Flux dependence of the transmission coefficient for the Fibonacci case
    at $E=0$ for a $6$th generation array. The $A$- and the $B$-type
    rings have, respectively, (a) $n_A=m_A=4$ and $n_B=m_B=6$ and (b)
    $n_A=m_A=4$ and $n_B=m_B=9$. The three panels in (a) highlight the
    periodicity at $\phi_A=25\phi_0$.}

\end{figure}

We now consider the transmission for a fixed electron energy $E$
as the flux through the rings is varied. As $\phi_B/\phi_A =
S_B/S_A =N_{B}^2/N_{A}^2$ the periodicity of the transmission
coefficient should be sensitive to $S_B/S_A$ independent of the
order in which the rings are arranged, i.e., periodic or
quasiperiodic. Let us work out a specific example. We consider
$E=0$. Let the $A$-ring have $2p$ atoms in each arm and the
$B$-ring have $2q$ atoms in each arm, $p$ and $q$ being integers.
Here, $S_B/S_A = (2q+1)^2/(2p+1)^2$. The dimer matrices {\boldmath
$P$}$_A$ and {\boldmath $P$}$_B$ commute and for rings arranged in
an $m$-th generation Fibonacci array the transmission coefficient
is given by
\begin{equation}
  \label{eq:tau}
  T(m) = 4 \frac{\tau_{pq}^{m}(\phi_A)}{\left[ \tau_{pq}^{m}(\phi_A) + 1 \right]^2}
\end{equation}
with
\begin{eqnarray}
\tau_{pq}^{m}(\phi_A)& = & 2^{2F_m}
\cos^{2F_{m-1}}\left(\pi\frac{\phi_A}{\phi_0}\right) \nonumber \\
& & \mbox{ }\times
\cos^{2F_{m-2}}\left[\pi\frac{(2q+1)^2\phi_A}{(2p+1)^2\phi_0}\right].
\end{eqnarray}
Here, $F_m=F_{m-1}+F_{m-2}$ is the Fibonacci number in the $m$th
generation, with $F_0=F_1=1$. The above formula reproduces the
result of the corresponding {\it even-even} ordered case.
Resonance ($T=1$) is achieved when $\tau_{pq}^{m}(\phi_A)=1$. Only
if $S_B=l S_A$, with integer $l$, the period of the transmission
coefficient remains $\phi_0$. This implies $(2q+1)=\sqrt{l}
(2p+1)$, which indicates that $\phi_0$-periodicity in $T$ cannot
be achieved for arbitrary combination of $p$ and $q$. If $S_B\neq
l S_A$, the transmission coefficient has a period equal to
$(2p+1)^2\phi_0$. In Fig.\ \ref{fig6}(a) we plot $T$ against
$\phi_A/\phi_0$ for a sixth-generation Fibonacci chain. The
$A$-ring has $4$ atoms and the $B$-ring has $5$ atoms in each arm,
respectively. The diagram shows a periodicity equal to $25\phi_0$.
In Fig.\ \ref{fig6}(b) a $\phi_0$-periodic behavior is shown with
the $A$-ring having $4$ atoms in each arm and the $B$-ring having
$9$ atoms in each arm. So, both $\phi_0$ and $(2p+1)^2 \phi_0$
periodicity can be present.

\subsection{A special case}

We now present an interesting feature that reveals the sensitivity
of the transmission coefficient on the specific choices of the
number of scatterers. We select $\phi_A=0$. The $A$-ring has just
one atom in each arm, while in the $B$-ring we have two atoms in
each arm. Then at $E=\sqrt{3}$, we get $J=0$, {\boldmath
$P$}$_B=-\openone$, and
\begin{equation}
  \mbox{\boldmath $P$}_A=
\left( \begin{array}{cc}
    - \sqrt{3}/2 & -1/2 \\
      1/2 & -\sqrt{3}/2
\end{array}\right) ,
\end{equation}
so that {\boldmath
  $P$}$_A^6=-\openone$. As a result, every $12$th generation of the
  Fibonacci sequence consists of identical strings
of {\boldmath $P$}$_A$'s starting with the second generation. The
transmission coefficients will consequently repeat every $12$th
generation. We thus have a case where even a quasiperiodic array
of rings has periodically repeating values of the transmittivity
as the system grows in size.  We believe that this feature is also
likely to be
present for non-zero flux values as well.  This aspect in under
further investigation.

\section{Summary}
\label{sec-conclusions}

We have studied the transmission spectra of mesoscopic rings in
the presence of a magnetic flux within a tight-binding formalism.
We address both periodic and quasiperiodic arrays of rings.  In
the spirit of a renormalization-group decimation method
\cite{DamPW89,LeaRS99}, we convert each ring into an effective
dimer, and attach a series of such dimers to semi-infinite perfect
leads to study their transmittance. In view of a possible
experimental realization we set the values of the on-site
potential and the nearest neighbor hopping integral in the leads
to be same as those in the bulk. We find that for both geometries
the transmission coefficient at fixed values of energy may exhibit
both types of periods, viz, $\phi_0$ and $\phi_0/2$. For the
periodic array of rings we find an interesting result: If a ring
contains an even number of atoms in one of its arms and an odd
number in the other arm, then in the absence of any magnetic flux
an array of such rings shows a power-law decay in the transmission
coefficient as the system grows in size. This has been shown
analytically for a special energy $E=0$. Apart from this, the
field-induced resonance in periodic arrays of rings has also been
discussed.
For the quasiperiodic Fibonacci array of rings of two different
sizes we find a drastic change in the distribution of allowed
energy values compared to the purely one-dimensional Fibonacci
chain \cite{KohKT83}. The magnetic field alters it further. The
trace maps and the Fibonacci invariant have been derived including
the magnetic field. The variation of the transmission coefficient
as functions of the external flux, as well as the energy of the
incident electron has been studied. The transmission coefficient
exhibits different periodicities depending on the relative areas
of the rings. The formulation has also been tested for other
aperiodic sequences such as the Thue-Morse sequence. The basic
nature of the periodic variation of the transmittivity as a
function of the magnetic flux remains the same.

\begin{acknowledgments}
  Stimulating discussions with I.\ Peschel are gratefully acknowledged.
  AC thanks the DAAD for financial support during his stay at
  Chemnitz University of Technology, and in particular the theory group
  in Chemnitz for their hospitality. RAR and MS gratefully acknowledge
  support by the DFG via SFB393.
\end{acknowledgments}



\end{document}